\documentclass[twocolumn,aps,showpacs,prl,groupedaddress]{revtex4}
\usepackage{epsfig,amssymb,amsmath}
\usepackage[hypertex,linkcolor=red]{hyperref}
\def\comment#1{}\def\labell#1{\label{#1}}
\def\section#1{{\par\em #1:--- }}
\def\togli#1{}

\begin{document}
\title{Closed timelike curves via post-selection: theory and
 experimental demonstration} \author{Seth Lloyd$^{1}$, Lorenzo
 Maccone$^{1}$, Raul Garcia-Patron$^{1}$, Vittorio Giovannetti$^{2}$,
 Yutaka Shikano$^{1,3}$, Stefano Pirandola${^1}$, Lee A.
 Rozema$^{4}$, Ardavan Darabi$^{4}$, Yasaman Soudagar$^{4,5}$, Lynden
 K.  Shalm$^{4}$, and Aephraim M.  Steinberg$^{4}$}
\affiliation{$^{1}$xQIT,Massachusetts Institute of
 Technology, 77 Mass Ave, Cambridge MA.\\
 $^2$NEST-CNR-INFM \& Scuola Normale Superiore, Piazza dei Cavalieri
 7, I-56126, Pisa, Italy. \\ $^3$Dep. Physics, Tokyo Institute of
 Technology, 2-12-1 Oh-Okayama, Meguro, Tokyo, 152-8551, Japan.\\
 $^4$CQIQC,IOS, Department of Physics, University of Toronto, Canada
 M5S 1A7.  \\$^5$Laboratoire des fibres optiques, \'Ecole
 Polytechnique de Montr\'eal, Eng. Phys. Dep., Montr\'eal, Canada.}

\begin{abstract}
 Closed timelike curves (CTCs) are trajectories in spacetime that
 effectively travel backwards in time: a test particle following a
 CTC can in principle interact with its former self in the past.
 CTCs appear in many solutions of Einstein's field equations and any
 future quantum version of general relativity will have to reconcile
 them with the requirements of quantum mechanics and of quantum field
 theory.  A widely accepted quantum theory of CTCs was proposed by
 Deutsch.  Here we explore an alternative quantum formulation of CTCs
 and show that it is physically inequivalent to Deutsch's.  Because
 it is based on combining quantum teleportation with post-selection,
 the predictions/retrodictions of our theory are experimentally
 testable: we report the results of an experiment demonstrating our
 theory's resolution of the well-known `grandfather
 paradox.'\end{abstract}
\pacs{03.67.-a,03.65.Ud,04.00.00,04.62.+v,04.60.-m} 
\maketitle
Although time travel is usually taken to be the stuff of science
fiction, it is not ruled out by scientific fact. Einstein's theory of
general relativity admits the possibility of closed timelike curves
(CTCs)~\cite{GODEL}, 
paths through spacetime which, if followed, allow a time traveller to
go back in time and interact with her own past.  The logical paradoxes
inherent in time travel make it hard to formulate self-consistent
physical theories of time
travel~\cite{DEU,Hawking,Deser,Politzer,THORNE}.  This paper proposes
an empirical self-consistency condition for closed timelike curves: we
demand that a generalized measurement made before a quantum system
enters a closed timelike curve yield the same statistics -- including
correlations with other measurements -- as would result if the same
measurement were made after the system exits from the curve.  That is,
the closed timelike curve behaves like an ideal, noiseless quantum
channel that displaces systems in time without affecting the
correlations with external systems. To satisfy this criterion without
introducing contradictions, we construct a theory of closed timelike
curves via quantum post-selection (P-CTCs).  The theory is based on
Bennett and Schumacher's suggestion~\cite{benschu} to describe time
travel in terms of quantum teleportation, and on the
Horowitz-Maldacena model for black hole
evaporation~\cite{HM}. 
We show that P-CTCs are consistent with path integral
approaches~\cite{politzer1,hartle}, but physically inequivalent to the
prevailing theory of closed timelike curves due to Deutsch~\cite{DEU}.
Moreover, because they are based on post-selection~\cite{SVE}, closed
timelike curves can be simulated experimentally. We present an
experimental realization of the grandfather paradox: the experiment
tests what happens when a photon is sent a few billionths of a second
back in time to try to `kill' its former self.

\begin{figure}[htb]
\begin{center}
\epsfxsize=.5\hsize\epsffile{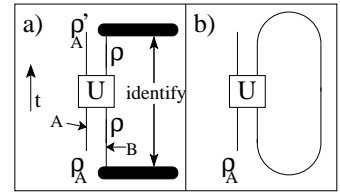} 
\end{center}
\caption{ a) Deutsch's quantum description of CTCs is based on the
 consistency condition of Eq.~(\ref{conscond}), where the unitary $U$
 describes an interaction between a chronology-respecting system $A$,
 initially in the state $\rho_A$, and a system $B$ in a CTC.  Deutsch
 demands that the state $\rho$ of $B$ at the input and output of $U$
 be equal, inducing a nonlinear transformation $\rho_A\to\rho'_A$.
 Time goes from bottom to top in this and in the following diagrams.
 b) P-CTC: post-selected quantum teleportation is employed as a
 description of the closed timelike
 curve. 
 The bottom curve $\bigcup$ represents the creation of a maximally
 entangled state of two systems and the upper curve $\bigcap$
 represents the projection onto the same state. 
\label{f:deu}}
\end{figure}

Deutsch's elegant quantum treatment of closed timelike
curves~\cite{DEU} provides a self-consistent resolution of the various
paradoxes of time travel by requiring simply that a system that enters
such a curve in a particular quantum state $\rho$, emerges in the past
in the same state (Fig.~\ref{f:deu}a) even after interacting with a
``chronology-respecting'' system in a state $\rho_A$ through a unitary
$U$. This translates into the consistency condition,
\begin{eqnarray}
\rho=\mbox{Tr}_A[U(\rho\otimes\rho_A)U^\dag]
\labell{conscond}\;.
\end{eqnarray}
A state $\rho$ that satisfies Eq.~(\ref{conscond}) always exists
because the above interaction is a completely positive map which
possesses at least one fixed point.

Here we propose an alternative consistency condition for CTCs: a
generalized measurement made on the state entering the curve should
yield the same results, including correlations with other
measurements, as would occur if the same measurement were made on the
state emerging from the curve. The CTC should behave like an ideal
quantum channel (even though, as we shall see, inside a CTC a proper
definition of state cannot be given). Deutsch's CTCs fail this
requirement, thus are physically inequivalent to our proposed
solution. To make the CTC behave like a quantum channel, we describe
it using quantum teleportation~\cite{teleport}, namely the perfect
transfer of an unknown quantum state $|\psi\rangle$ between two
parties (Alice and Bob) using a shared entangled state, the
transmission of classical information, and a unitary transformation
$V$ on Bob's side.
A curious feature of teleportation is that, whenever Alice's Bell
measurement gives the same result it would when measuring the initial
shared state, then Bob's unitary $V$ is the identity.  In this case,
Bob does not have to perform any transformation to obtain Alice's
state $|\psi\rangle$: in some sense, Bob possesses the unknown state
even before Alice implements the teleportation!  Causality is not
violated because Bob cannot foresee Alice's measurement result, which
is completely random. However, if we could
pick out only the proper result with probability one, the resulting
`projective' teleportation would allow information to propagate along
spacelike intervals, to escape from black
holes~\cite{HM}, 
or to travel backwards in time along a closed timelike curve.  We call
this mechanism a projective or post-selected CTC, or P-CTC. 

The P-CTC (see Fig.~\ref{f:deu}b) starts from two systems prepared in
a maximally entangled state ``$\bigcup$'', and ends by projecting them
into the same state ``$\bigcap$''. If non-zero, the probability
amplitude of the final state is renormalized to one, a nonlinear
process. If zero, the event cannot happen: our mechanism embodies in a
natural way the Novikov principle~\cite{novikov} that only {\it
 logically self-consistent} sequences of events occur in the
universe.  Because they rely on post-selection, P-CTCs share some
properties with the weak value interpretation of quantum
mechanics~\cite{weak}, notably that there is no unique way to assign a
definite state to the system in a CTC at a definite time.  This is not
surprising due to the cyclic nature of time there. Moreover,
Hartle~\cite{hartle} showed that quantum mechanics on closed timelike
curves is non-unitary (indeed, it allows cloning) and requires events
in the future to affect the past.  He noted that the Hilbert space
formalism for quantum mechanics might be inadequate to capture the
behavior of closed timelike curves, and suggested a path integral
approach instead.  In future work we will show that, in
contrast~\cite{politzer1} to Deutsch's, P-CTCs are consistent with the
``traditional'' path-integral approaches to CTCs
(e.g.~see~\cite{politzer1,Politzer,hartle,friedman}): we use the
normal path-integral self-consistency requirement that the classical
paths that make up the path integral have the same values of all
variables (e.g.~$x$ and $p$) when they exit the CTC as when they
enter.  For example, our approach coincides with
Politzer's~\cite{politzer1} path-integral treatment of fermions.
However, P-CTCs can also be described in Hilbert space, showing that
this approach can be reconciled with path integrals when
post-selection is allowed.

\begin{figure}[htb]
\begin{center}
\epsfxsize=.55\hsize\epsffile{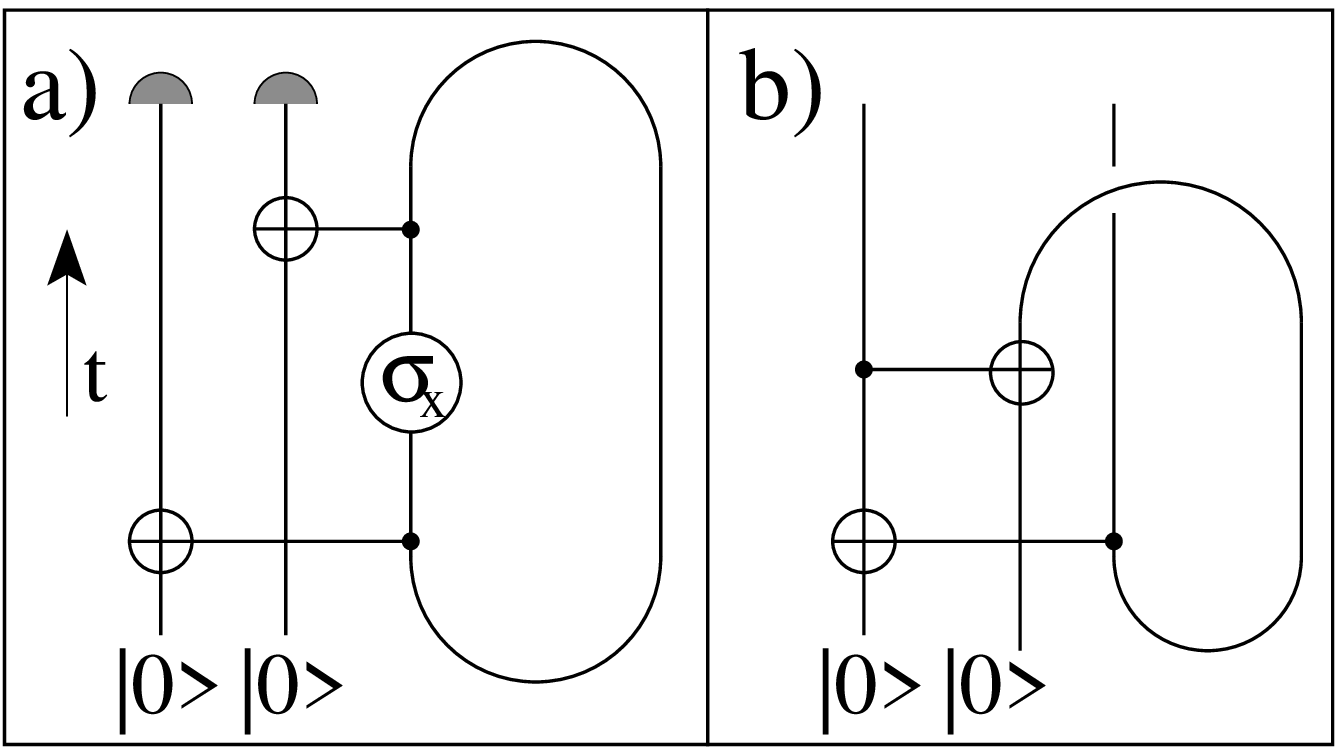} 
\end{center}
\caption{ a) Grandfather paradox circuit.  
 If we take $1$ to represent `time-traveler exists,' and $0$ to
 represent `she doesn't exist,' then the NOT ($\sigma_x$) operation
 implies that if she exists, then she `kills her grandfather' and
 ceases to exist; conversely, if she doesn't exist, then she fails to
 kill her grandfather and so she exists.  The difference between
 Deutsch's CTCs and our P-CTCs is revealed by monitoring the
 time-traveler with two controlled-NOTs (CNOT): the two controlled
 bits are measured to determine the value of the time-traveling bit
 before and after the $\sigma_x$.  Opposite values mean she has
 killed her grandfather; same values mean she has failed. Using
 Deutsch's CTCs, she always succeeds; using P-CTCs she always fails.
 b) Unproved theorem paradox circuit. The time-traveler obtains a bit
 of information from the future via the upper CNOT.  She then takes
 it back in time and deposits a copy an earlier time in the same
 location from which she obtained it (rather, will obtain it), via
 the lower CNOT.  Because the circuit is unbiased as to the value of
 the `proof' bit, it automatically assigns that bit a completely
 mixed value, since it is maximally entangled with the one emerging
 from the post-selected CTC.
\label{f:grandfather}}
\end{figure}

We now analyze how P-CTC deal with time travel paradoxes. In the
grandfather paradox, for example, the time traveller goes back in time
and kills her grandfather, so she cannot be born and cannot kill her
grandfather: a logical contradiction.
This paradox can be implemented through a quantum circuit where a
`living' qubit (i.e., a bit in the state 1), goes back in time and
tries to `kill' itself, i.e., flip to the state 0, see
Fig.~\ref{f:grandfather}a. There are many possible variants: but any
circuit in which travels back in time and gives rise to intrinsic
logical self-contradiction is an embodiment of the grandfather
paradox.  Deutsch's consistency condition (\ref{conscond}) requires
that the state
is 
$\rho=(|0\rangle\langle 0|+|1\rangle\langle
1|)/2$, the only fixed point of the corresponding map. 
Note that if the CNOT before the bit flip measures a 0 then the CNOT
afterwards measures a 1, and {\it vice versa}: the time traveller
really manages to kill her grandfather!  So far, so good.  The strange
aspect of Deutsch's solution comes when one attempts to follow the
state of the time-traveller through the CTC. To preserve
self-consistency, the $1$ component (time traveller alive) that enters
the loop emerges as the $0$ component (time traveller dead), while the
$0$ component (time traveller dead) that enters the loop emerges as
the $1$ component (time traveller alive).  Thus, the CTC preserves the
overall mixed state, but not the identity of the components:
projective measurements at the input and output yield different
results.  

P-CTCs give a different resolution of the grandfather paradox: the
probability amplitude of the projection onto the final entangled state
$\bigcap$ is always null, namely this event (and all logically
contradictory ones) cannot happen. In any real-world situations, the
$\sigma_x$ transformation in not perfect.  Then, replacing $\sigma_x$
with $e^{-i\theta\sigma_x}=\cos\tfrac\theta
2\openone-i\sin\tfrac\theta 2\sigma_x$ (with $\theta\simeq\pi$), the
non-linear post-selection amplifies fluctuations of $\theta$ away from
$\pi$. This eliminates the histories plagued by the paradox and
retains only the self-consistent histories in which the time-traveler
fails to kill her grandfather (the unitary in the curve is $\openone$
instead of $\sigma_x$), and the two output qubits have equal value:
P-CTC fulfill our self-consistency condition. In other words, no
matter how hard the time-traveler tries, she finds her grandfather a
tough guy to kill.

\begin{figure*}[htb]
\begin{center}
\epsfxsize=.7\hsize\epsffile{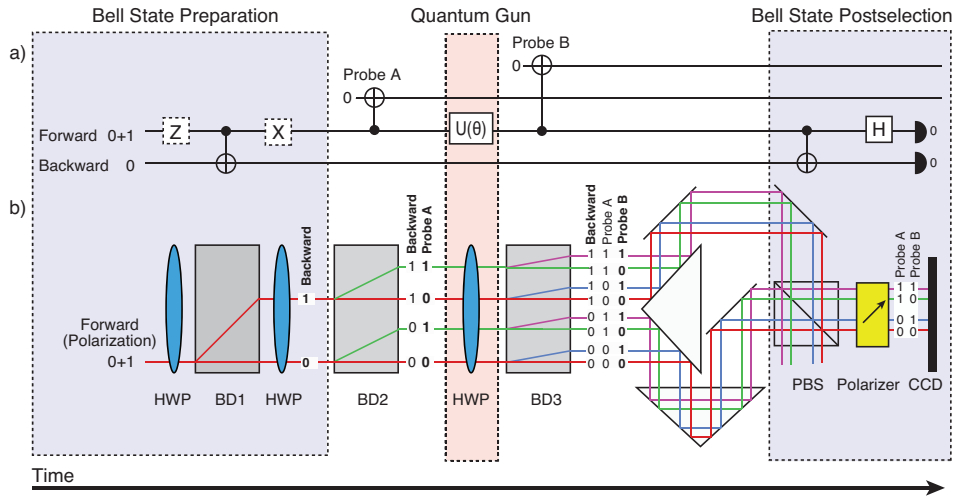}
\end{center}
\caption{Experiment to illustrate the P-CTC predictions of the
  grandfather paradox.  a) Diagram of the quantum circuit.  Using a
  CNOT gate sandwiched between optional Z and X gates, it is possible
  to prepare all of the maximally entangled Bell states.  The Bell
  state measurement is implemented using a CNOT and a Hadamard.  Each
  of the probe qubits is coupled to the forward qubit via a CNOT gate.
  b) Diagram of experimental apparatus.  The polarization and path
  degrees of freedom of single photons from a quantum dot are
  entangled via a calcite polarization-dependent beam displacer (BD1),
  implementing the CNOT.  Half-wave plates (HWP) before and after BD1
  implement the optional $Z$ and $X$ gates.  The state
  $|\phi^{+}\rangle$ is created by setting the angle of both HWPs to
  zero.  To complete the teleportation circuit, the post-selection
  onto $|\phi^{+}\rangle$ is carried out by first recombining the path
  degrees of freedom on a polarizing beamsplitter (performing a CNOT
  gate between path and polarization) and then passing the photons
  through a calcite polarizer set to $45$ degrees and detecting them
  on a cooled CCD. A rotatable HWP acts as a quantum gun, implementing
  the unitary $U(\theta)=e^{-i\theta\sigma_x}$.  Removable calcite
  beam displacers (BD2 and BD3) couple the polarization qubit to two
  probe qubits encoded in additional spatial degrees of freedom.  When
  the beam displacers are inserted in the setup, four spots on the CCD
  correspond to the probe states $11$, $10$, $01$, and $00$.
\label{f:experiment}}
\end{figure*}
Because P-CTCs are based on post-selected teleportation, their
predictions can be experimentally demonstrated.  To experimentally
demonstrate the grandfather paradox, we store two qubits in a single
photon: one in the polarization degree of freedom, which represents
the forward-travelling qubit, and one in a path degree of freedom
representing the backward travelling qubit as shown in Fig
\ref{f:experiment}.  Our single photons, with a wavelength of $941.7$
nm, are coupled into a single-mode fiber from an InGaAs/GaAs quantum
dot cooled to $21.5 K$ by liquid Helium ~\cite{Mirin} and sent to the
circuit.  Using a Hanbury-Brown-Twiss interferometer, the $g^{(2)}(0)$
of the quantum dot emission was measured to be $0.29 \pm 0.01$,
confirming the single-photon character of the source.  At the start of
the circuit, ($\bigcup$) we entangle the path and polarization qubits
using a beam displacer (BD1), generating the Bell state
$|\phi^{+}\rangle=(1/\sqrt{2})(|00\rangle+|11\rangle)$.  To close the
CTC ($\bigcap$), we perform a Bell state measurement and post-select
on cases where $|\phi^{+}\rangle$ is detected.  The Bell state
measurement consists of a CNOT with polarization (forward traveller)
acting on path (backward traveller), followed by post-selection on the
now-disentangled qubits.  The CNOT is implemented by a polarizing beam
splitter that flips the backward-travelling (path) qubit conditioned
on the value of the forward-travelling (polarization) qubit.  We then
post-select on photons exiting the appropriate spatial port using a
polarizer at $45^\circ$ and an Andor iDus CCD camera cooled to $188
K$.  Within the loop, we implement a ``quantum gun''
$e^{i\theta\sigma_x}$ with a wave plate that rotates the polarization
by an angle $\theta/2$.  The accuracy of the quantum gun can be varied
from $\theta=\pi$ (the photon ``kills'' its past self) to $\theta=0$
(the photon always ``misses'' and survives).

\begin{figure*}[h!]
\begin{center}
\epsfxsize=.6\hsize\epsffile{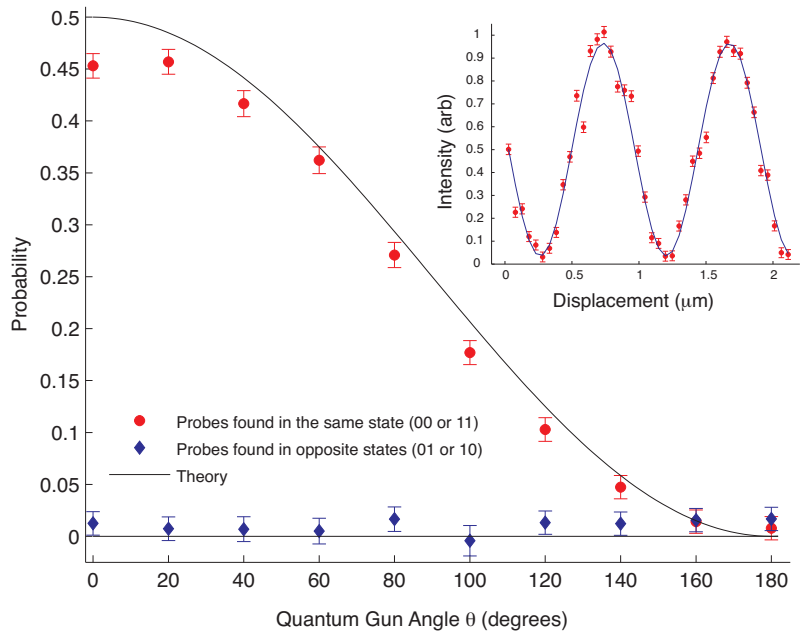}
\end{center}
\caption{ Probability that time travel succeeds and the probes are 
found in the same state (red circles) or in opposite states 
(blue diamonds).  When the quantum gun ``misfires'', the 
polarization qubit is not flipped and the probe qubits are 
found in either the $00$ or $11$ state.  As the accuracy $\theta$
of the quantum gun increases from 
$0$ to $\pi$, the probability that the teleportation 
succeeds decreases.  When the quantum gun ``kills'' the photon 
(flips the polarization qubit), the probes record opposite values 
($01$ or $10$).  The probability that the probe qubits are found 
in either the $10$ or $01$ state is $0.01 \pm 0.04$, indicating 
that the photons never succeed in travelling back in time and 
killing their former selves.  Solid curves correspond to theoretical 
predictions. The discrepancy between theory and experiment when the 
probes are found in the same state is due to a $1.1 \pm 0.1^{\circ}$ 
mismatch between polarizers used in the state creation and measurement 
portions of the teleportation circuit.  Data were collected for 6 seconds 
at each point.  The error bars are due to photon counting statistics and 
background fluctuations from the cooled CCD.  Inset: the teleportation loop 
constitutes a polarization interferometer.  Its visibility was
measured 
as $93 \pm 3 \%$ by varying the phase (path-length 
difference) between the two paths, converting $|\phi^{+}\rangle$ 
to $(|00\rangle+e^{i\phi}|11\rangle)/\sqrt{2}$ before 
post-selecting on $|\phi^{+}\rangle$.  
\label{f:results}}
\end{figure*}

The teleportation circuit forms a polarization interferometer whose visibility was measured to be $93 \pm 3 \%$ (see the inset in Fig.~\ref{f:results}).
To verify the operation of the teleportation circuit, all four Bell states $|\phi^{\pm}\rangle,|\psi^{\pm}\rangle$ were prepared and sent to the measurement apparatus: post-selection on $|\phi^{+}\rangle$ behaved as expected yielding success probabilities of  $0.96 \pm 0.08$, $0.10 \pm 0.11$, $0.02 \pm 0.05$, and $0.02 \pm 0.05$   for $|\phi^{+}\rangle$, $|\phi^{-}\rangle$, $|\psi^{+}\rangle$, and $|\psi^{-}\rangle$ inputs respectively.  After verifying the operation of the teleportation circuit, calcite beam displacers were inserted (BD2 and BD3), coupling the polarization qubit to two probe qubits encoded in additional path degrees of freedom of the photon.  These probe qubits measure the state of the polarization qubit before and after the quantum gun is ``fired''.  When the post-selection succeeds (i.e. the time travel occurs), the state of the probe qubits is measured.  If the two probe qubits are in agreement (00 or 11) the quantum gun has failed to flip the polarization and the photon ``survives''.  If the two probe qubits disagree, the photon has ``killed'' its past self.  
 
 The state of the probe qubits, conditioned on the time travel succeeding, was measured for different values of $\theta$, see Fig.~\ref{f:results}.  The probe qubits are never found in the states 01 or 10: time travel succeeds only when the quantum gun misfires, leaving the polarization unchanged and the probe qubits in 00 or 11.  Our suicidal photons obey the Novikov principle and never succeed in travelling back in time and killing their former
selves.   The required nonlinearity is due to post-selection here: no CTCs nor any evidence of the nonlinear signature of a P-CTC has ever been observed in nature up to now.


Unlike Deutsch's CTCs, our P-CTCs always send pure states to pure
states: they do not create entropy.  As a result, P-CTCs provide a
distinct resolution to Deutsch's unproved theorem paradox, in which
the time traveller reveals the proof of a theorem to a mathematician,
who includes it in the same book from which the traveller has learned
it (rather, {\em will} learn it). How did the proof come into
existence?  Deutsch adds an additional maximum entropy postulate to
eliminate this paradox.  By contrast, post-selected CTCs automatically
solve it as shown in Fig.~\ref{f:grandfather}b through entanglement:
because the circuit has no bias to one proof or another, the CTC
creates an unbiased mixture of all possible `proofs.'

A user that has access to a closed timelike curve might be able to
perform computations very efficiently: for pure state inputs,
Deutsch's CTCs permit the efficient solution of all problems in
PSPACE~\cite{aar1,Brun} (i.e.~all problems that can be solved with
polynomial space resources).  However, Bennett {\it et al.} argued
that this may be useless for computation, because CTCs decorrelate the
outputs of the computation from its inputs stored
elsewhere~\cite{BEN}.  In contrast, Aaronson's results on the power of
post-selection in quantum computing imply that the P-CTCs considered
here permit the efficient solution of problems in PP~\cite{aar2}
(i.e.~problems that a probabilistic polynomial Turing machine accepts
with probability $\tfrac 12$ if and only if the answer is ``yes.'')
Although PP is a putatively less powerful class than PSPACE, P-CTCs
are computationally very powerful: they do not decorrelate the inputs
from the outputs and can efficiently solve NP-complete problems.
Indeed, it is easy to see that P-CTCs can perform any computation a
circuit of depth one.

We thank C.~Bennett and D.~Deutsch for discussions and suggestions.
This work was supported by the W.M. Keck foundation, Jeffrey Epstein,
NSF, ONR, DARPA, JSPS, NSERC, Quantum Works, CIFAR.  R.~Mirin and
M.~Stevens from the Optoelectronics Division at NIST provided the
quantum dots for the experiment.


\begin{references}
\bibitem{GODEL} K. G\"{o}del, Rev.  Mod. Phys. {\bf 21}, 447 (1949);
 M.S. Morris, K.S. Thorne, U. Yurtsever, Phys. Rev.  Lett.  {\bf 61},
 1446 (1988);
B. Carter, Phys. Rev. {\bf 174}, 1559 (1968);
W.J. van Stockum, Proc. Roy. Soc. A {\bf 57}, 135 (1937);
J.R. Gott, Phys. Rev. Lett. {\bf 66}, 1126 (1991).

\bibitem{DEU}D. Deutsch, Phys. Rev. D {\bf 44}, 3197 (1991).

\bibitem{Hawking} S. Hawking, Phys. Rev. D {\bf 46}, 603 (1992).

\bibitem{Deser} S. Deser, R. Jackiw, G. 't Hooft, Phys. Rev.  Lett.
 {\bf 68}, 267 (1992).

\bibitem{Politzer} H.D. Politzer, Phys. Rev. D {\bf 46}, 4470 (1992).

\bibitem{THORNE} S.-W. Kim, K.S. Thorne, Phys. Rev. D {\bf 43}, 3929
 (1991).

\bibitem{benschu} C.H. Bennett, talk at QUPON, Wien, May 2005;
 http://www.research.ibm.com/people/b/bennetc/.

\bibitem{HM} G.T. Horowitz, J. Maldacena, JHEP {\bf 02}, 8 (2004);
U. Yurtsever, G. Hockney, Class. Quant.  Grav.
 {\bf 22}, 295 (2005);
D. Gottesman, J. Preskill, JHEP {\bf 03}, 26 (2004);
S. Lloyd, Phys. Rev. Lett. {\bf 96}, 061302 (2006).

\bibitem{politzer1} H.D. Politzer, Phys. Rev. D {\bf 49,} 3981 (1994).

\bibitem{hartle} J.B. Hartle, Phys. Rev. D {\bf 49}, 6543 (1994).

\bibitem{SVE} G. Svetlichny, arXiv:0902.4898 (2009).

\bibitem{teleport} C.H. Bennett, G. Brassard, C. Cr\'epeau, R. Jozsa,
 A. Peres, W.K. Wootters, Phys. Rev.  Lett. {\bf 70}, 1895 (1993).

\bibitem{novikov} J. Friedman, {\it et al.} 
 Phys. Rev. D {\bf 42}, 1915 (1990).

\bibitem{weak} Y. Aharonov, D.Z. Albert, L. Vaidman, Phys. Rev. Lett.
 {\bf 60}, 1351 (1988).

\bibitem{friedman}J.L. Friedman, N.J. Papastamatiou, J.Z. Simon, Phys.
 Rev. D {\bf 46}, 4456 (1992).


\bibitem{Mirin}R.P.  Mirin, App. Phys. Lett. {\bf 84}, 1260 (2004).
\bibitem{aar1}S. Aaronson, J. Watrous, Proc. Roy. Soc. A {\bf 465},
631 (2009); arXiv:0808.2669v1.

\bibitem{Brun} T.A. Brun, Found. Phys. Lett. {\bf 16}, 245 (2003);
 T.A. Brun, J. Harrington, M.M. Wilde, Phys.  Rev.  Lett. {\bf 102},
 210402 (2009).
\bibitem{BEN} C.H. Bennett, D. Leung, G. Smith, J.A. Smolin, Phys.
 Rev. Lett. {\bf 103}, 170502 (2009).

\bibitem{aar2}S. Aaronson, Proc. Roy. Soc. A {\bf 461}, 3473 (2005).

\end{references}
\end{document}